\documentclass[aps,prx,floatfix,twocolumn,superscriptaddress]{revtex4-1}
\usepackage{amsmath,amssymb}
\usepackage{graphicx}% Include figure files
\usepackage{epsfig}% Include figure files
\usepackage{dcolumn}% Align table columns on decimal point
\usepackage{bm}% bold math
\usepackage{times}
\usepackage{epstopdf}
\usepackage{subfig}
\usepackage{paralist}
\usepackage{color}

\def\<{\langle}
\def\>{\rangle}
\def\(({\left(}
\def\)){\right)}
\def\[[{\left[}
\def\]]{\right]}

\begin{document}

\title{Ensemble renormalization group for disordered systems}

\author{Maria Chiara Angelini}
\affiliation{Dipartimento di Fisica, Universit\`{a} ``La Sapienza'', P.le A. Moro 5, I-00185 Roma, Italy}

\author{Giorgio Parisi}
\affiliation{Dipartimento di Fisica, INFN -- Sezione di Roma 1, CNR -- IPCF UOS Roma, Universit\`{a} ``La Sapienza'', P.le A. Moro 5, I-00185 Roma, Italy}

\author{Federico Ricci-Tersenghi}
\affiliation{Dipartimento di Fisica, INFN -- Sezione di Roma 1, CNR -- IPCF UOS Roma, Universit\`{a} ``La Sapienza'', P.le A. Moro 5, I-00185 Roma, Italy}

\begin{abstract}
  We propose and study a Renormalization Group transformation that can
  be used also for models with strong quenched disorder, like spin
  glasses.  The method is based on a mapping between disorder
  distributions, chosen such as to keep some physical properties
  (e.g. the ratio of correlations averaged over the ensemble)
  invariant under the transformation. We validate this Ensemble
  Renormalization Group (ERG) by applying it to the hierarchical model
  (both the diluted ferromagnetic version and the spin glass version),
  finding results in agreement with Monte Carlo simulations.
\end{abstract}

\maketitle
\section{Introduction}
The renormalization group (RG) is a fundamental tool in theoretical
physics \cite{bookAmitMM}. It allows to characterize phase transitions
and critical phenomena, by computing critical exponents and
universality classes.  The real space RG can be viewed as a decimation
procedure that takes a system made of $N$ dynamical variables and
reduces it to a smaller system, in a way which preserves, or scales
appropriately, some important physical observables.  Such a decimation
induces a RG transformation on the system couplings, and the study of
such a transformation allows one to identify critical points and
critical exponents.

Real space RG transformations have been studied in great detail for
homogeneous models \cite{bookAmitMM}, but much less is known for
disordered models, that contain quenched randomness in the Hamiltonian
(either random fields and/or random couplings). This is specially true
for strongly frustrated models, like spin glasses (SG), for which a
satisfying RG transformation is still lacking.  For example, for the
Edwards-Anderson SG model \cite{EA} on a $D$-dimensional lattice all
the attempts to develop a field theory by performing an
$\epsilon$-expansion around the upper critical dimension $D_u=6$ have
proved to be very complicated \cite{RGBray,temesv}.  These studies
have led to the discovery of fixed points different from the
mean-field (MF) ones, however the implications of that are not
completely clear. In particular the existence of replica symmetry
breaking (RSB) fixed points in the non-mean field region $D < D_u$ has
been shown only very recently \cite{temGiorgio} and estimates of
critical exponents in $D=3$ are still not reliable.

In this framework the development of a (semi-)analytical real space RG
for disordered models would be very welcome.  The outcome of such a RG
transformation could be well compared with Monte Carlo (MC)
simulations, that provide accurate estimates of critical temperatures
and critical exponents for disordered models (at a much higher
computational cost).

Since the first developments of RG transformations for models with
quenched disorder it was clear that one has to deal with distributions
of couplings \cite{Lubensky75,YoungStinchcombe75}.
But the RG transformation for an entire coupling distribution is highly
non-trivial.
Previous attempts of developing a real
space RG for disordered systems \cite{RSRGCastellana} focused on
transformations mapping a single sample of size $N$ to a smaller
system (without loss of generality we can set the size of the smaller
system to $N/2$).  In formulae, we can write the mapping $\{J_{ij}\}
\to \{J'_{ij}\}$ as the one solving a set of equations like
\begin{equation}
\< O_k(\{J_{ij}\}) \> = \< O'_k(\{J'_{ij}\}) \>\;,
\label{oldRG}
\end{equation}
where angular brackets are thermal averages with respect to the
Gibbs-Boltzmann distribution, primed quantities refer to the smaller
system and the number of observables $O_k$ is enough to determine the
new couplings $\{J'_{ij}\}$.  Given an ensemble of systems of size
$N$, the above transformation can be applied to each of them in order
to obtain an ensemble of systems of size $N/2$.

However we believe
that such a mapping is suboptimal for models with quenched disorder
and a better RG transformation should consider explicitly the average
over the quenched disorder (as was done e.g.\ in
Ref.~\cite{ParisiPetronzio}). What we are proposing is a mapping
between probability distributions of couplings $P(J_{ij}) \to
P'(J'_{ij})$ such that the following equations hold
\begin{equation}
\overline{\< O_k(\{J_{ij}\}) \>} = \overline{\< O'_k(\{J'_{ij}\}) \>}\;.
\label{ourRG}
\end{equation}
The overbar represents the average over the quenched disorder
(i.e.\ the couplings in the present case).  The rationale beyond this
choice is that in models with strong disorder (like SG)
sample-to-sample fluctuations may dominate thermal ones.

It is worth noticing that standard RG transformations working
on coupling distributions typically use Eq.~(\ref{oldRG}).
For example, the simplest RG transformation for the
bond percolation problem in 2 dimensions ($p'=2p^2-p^4$) or even the
more complicated one of Ref.~\cite{YoungStinchcombe75} can be viewed
as single sample transformations, where $p$ is the bond density by
which the larger samples are generated and $p'$ is the bond density
of the renormalized samples (one-by-one).
Instead, the new approach in Eq.~(\ref{ourRG}) would prescribe to
compute some physical quantity (e.g.\ a correlation or a probability
of being connected) in the ensemble of larger systems with
bond density $p$, and to repeat the same computation in the ensemble
of smaller systems with several bond densities in order to find the
best value $p'$ such that the physical observables match in the two
ensembles.
The new approach we are proposing resembles in some way the finite
size scaling analysis which is performed on Monte Carlo data measured
in systems of different sizes.
Obviously, for models where the RG transformation is exact (like
e.g.\ the diamond hierarchical lattices \cite{McKay82}) the old and
the new approaches provide the same answer.

Two simple examples may help elucidating even more the limits of the RG
transformation working sample by sample, Eq.(\ref{oldRG}), and thus
justify the use of the one in Eq.(\ref{ourRG}), that we will call
Ensemble RG (ERG). In a diluted ferromagnet with long range interactions,
where couplings are
positive with probability $p$ and null with probability $1-p$, a
single step of the decimation procedure induced by Eq.(\ref{oldRG})
typically generates all non-zero couplings (i.e. $p=1$). And this is
clearly not very useful if one is willing to follow the RG flow in the
$p-T$ plane. Moreover, in frustrated models the decimated system is
typically much less frustrated than the original one: the extreme case
is the transformation of a 4-spins system in a 2-spins system, being
the latter unfrustrated for any coupling choice! This tendency to
reduce frustration makes the RG using Eq.(\ref{oldRG}) clearly unfit
to describe SG fixed points.

In principle, our ERG scheme can be applied to any disordered system.
We choose here to apply it to the hierarchical model (HM), which is a
particular one-dimensional long range model, whose Hamiltonian for
$N=2^n$ spins can be constructed iteratively in the following way
\cite{Dyson}:
\begin{multline}
H_n(s_1,...,s_{2^n}) = H_{n-1}(s_1,...,s_{2^{n-1}}) +\\
+ H_{n-1}(s_{2^{n-1}+1},...,s_{2^n}) + c^n \sum_{i<j=1}^{2^n} J_{ij}\,s_i\,s_j\;.
\label{eq:Hindip}
\end{multline}
In practice $H_n$ is the sum of interactions at $n$ different levels.
We have studied three versions of this model: the ferromagnet (FM),
where $J_{ij}=1$; the diluted ferromagnet (DFM), where a random
fraction $1-p$ of FM couplings are set to zero; and the SG version
\cite{HEAM}, with Gaussian couplings $P(J) \propto e^{-J^2/2}$.

There are many reasons to test a new RG transformation on the
HM. First, by properly tuning the topological factor $c$ that controls
how fast the couplings intensity decays with distance, the HM can
emulate a $D$-dimensional short range (SR) model: $c \simeq
2^{-1-\frac{2}{D}}$ for DFM and $c\simeq2^{(-1-\frac{2}{D})/2}$ for SG
\cite{cDrelations}.  These relations are exact around the upper
critical dimensions because the long range HM and the SR
$D$-dimensional model have the same field theory at leading order.  In
order to have a phase transition at a finite temperature, the $c$
parameter must satisfy
$c^{FM}_L=\frac{1}{4}<c<\frac{1}{2}=c^{FM}_{\infty}$ for the DFM and
$c^{SG}_L=\frac{1}{2}<c<\frac{1}{\sqrt{2}}=c^{SG}_{\infty}$ for the
SG.  Lower bound values $c_L$ correspond to lower critical dimensions
(and thus $T_c=0$), while when $c=c_{\infty}$ the energy is no longer
extensive (thus $T_c=\infty$).  For $c>2^{-\frac{3}{2}}=c^{FM}_U$ in
DFM and $c>2^{-\frac{2}{3}}=c^{SG}_U$ in SG, the model shows mean
field critical properties (like for $D>D_U$ in SR models).  So, tuning
a single parameter in the HM, we can move from the MF region to a
non-MF one.

The second reason to choose the HM, is that if the system is decimated
by a standard block-spin transformation, the new Hamiltonian does not
contain any multi-spin term (at variance to what happen on finite
dimensional lattices \cite{bookAmitMM}).  So, considering only
pairwise interactions in the RG is not an approximation for the HM.
Moreover the FM version can be exactly solved in a time growing only
polynomially with $N$, since the probability distribution of the
magnetization satisfies
\begin{equation}
p_n(m) \propto e^{\beta c^n m^2}\!\sum_{m_L,m_R} p_{n-1}(m_L)\,p_{n-1}(m_R)\,\delta_{m_L+m_R,m}
\label{eq:Pm}
\end{equation}
where $m_L$ and $m_R$ are the magnetizations of the half systems.
It is worth stressing that Eq.~(\ref{eq:Pm}) is no more valid for the
SG version of the HM, which does not admit a polynomial time solution.
In Ref.~\cite{eps_exp}, a generalization of Eq.~(\ref{eq:Pm}) has been
proposed to describe the distribution of the overlap $q$ in the SG
version of the HM; however we will not make use of this relation, due
to the difficulties in its treatment.

\begin{figure}[t]
\begin{center}	
\includegraphics[width=.7\columnwidth]{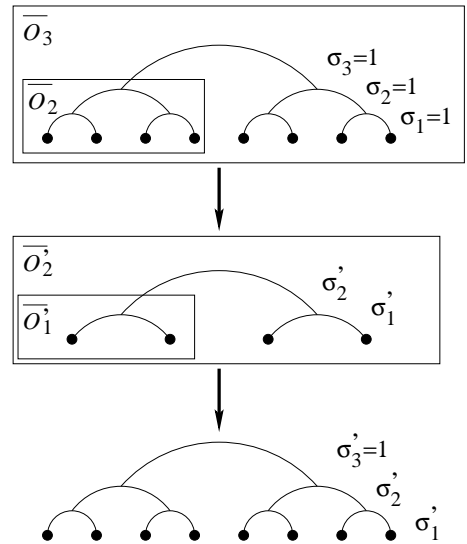}
\end{center}
\caption{\label{reduction} Schematic representation of the ERG for a
  SG HM with $n=3$. Variances ($\sigma_k$) and renormalized variances
  ($\sigma'_k$) at each level $k$ are indicated.}
\end{figure}

\section{Ensemble renormalization group for HM}

We describe now in detail how to apply the ERG to the HM.  We assume
couplings to remain independent during the RG, but we allow couplings
to have a different probability distribution $P_k(J)$ (or a different
value $J_k$ in the FM version) at each level $k\in\{1,2,...,n\}$: in
the original HM all couplings have the same probability law, but we
have seen that the RG iteration produces different couplings at
different levels.  In the analyzed versions, each coupling
distribution is parametrized by $K \in \{1,2\}$ few numbers (that is
the mean for FM, the variance for SG, the fraction of non-zero
couplings and the mean for DFM), otherwise the search for a solution
to Eq.(\ref{ourRG}) would become too difficult.  We start from an
ensemble of systems with $n$ levels that we want to reduce to an
``equivalent'' ensemble of smaller systems of $n-1$ levels. The whole
procedure is described in Fig.~\ref{reduction} for a SG HM with $n=3$.
\begin{compactenum}
\item First we compute $(n-1)K$ observables $\overline{\langle O_j
    \rangle}$, $j\in \{K+1,\ldots,Kn\}$ in the larger systems
  extracted from the original coupling distribution (in the FM the
  overbar can be omitted).
\item \label{step2}Then we identify the new distributions of couplings
  $P'_k(J')$, $k\in\{1,\ldots,n-1\}$ for the ensemble of smaller
  systems, i.e., we determine the new $(n-1)K$ parameters of the $P'$
  distributions, by requiring that $\overline{\langle O'_i
    \rangle}_{\bm{P'}}=\overline{\langle O_{i+K} \rangle}_{\bm{P}}$
  for any $i \in \{1,2,...,(n-1)K\}$.
\item Finally we build a new ensemble of systems of the original size.
  They are constructed joining with random couplings extracted from
  the original distribution $P_n(J)$ two smaller systems with
  couplings extracted from $P'_k(J')$, $k\in\{1,2,...,(n-1)\}$ found
  at step \ref{step2}.
\end{compactenum}
Primed quantities refer to the smaller systems.  In this case the
parameters to be determined are the variances of the Gaussian
distributions at each level.  The first two steps are the true
renormalization steps, while the latter is required to obtain a final
system size, that will allow us to iterate the method, until
convergence.  In the disordered versions, the assumption that the
renormalized coupling distributions are of the same type as the
original ones is an approximation.  However it can be easily improved
adding extra terms in the distributions.  Thermal averages are
computed exactly: this is easy to do in the FM, thanks to
Eq. (\ref{eq:Pm}), while in the DFM and in the SG we do it by
exhaustive enumeration, thus limiting us to a small number of levels
in the disordered cases.  The average over the disorder is not exact,
but taken over $\sim10^5$ samples.  Step 2 is actually accomplished by
minimizing $\sum_i (\overline{\langle O'_i \rangle} -
\overline{\langle O_{i+K} \rangle})^2$, and we have checked that the
reached minimum is always very close to zero.  Since couplings
distributions are different at each level, we do not see any better
option than extracting the new couplings at level $n$ in step 3 from
the original distribution $P_n(J)$.

The FM version of the HM, for which exact RG equations can be written,
is a benchmark for our numerical implementation of the ERG.  For
$c=2^{-\frac{5}{3}}$, that corresponds to $D\simeq3$, the critical
temperature is $T_c=0.848154717$ \cite{Meurice}, and the critical
exponents are $\eta=4/3$, $\gamma=1.299140730159(1)$ \cite{critexp},
leading to $\nu=1.948711095$ using the scaling relations.  To compare
this critical exponent $\nu$ with the one for a SR 3D Ising FM, it can
be shown \cite{sigmad} that the following equation should be used with
a proper $c$--$D$ relation
\begin{equation}
\nu_{LR}(c)=D\nu_{SR}(D)\;.
\end{equation}
In the present case $\nu_{LR}(c=2^{-5/3})/3=0.649570365$ is close to
$\nu_{SR}(3)=0.6301(4)$ \cite{Pelissetto}.

The observables $O_k$ that we use in the RG equations are the
correlation of the magnetization at level $k+1$, normalized by those
at level $k$, with $k\in\{1,\ldots,n-1\}$:
\begin{equation*}
\langle O_k \rangle = \frac{\langle m_{L_k} m_{R_k}\rangle}{\langle m_{L_k} m_{L_k}\rangle}\;. 
\end{equation*}
The denominator is needed to reduce finite size effects and to ensure
that a solution to the RG equations always exists (this is not true in
general for other observables, like the magnetization).

\begin{figure}[t]
\begin{center}	
\subfloat{\includegraphics[width=0.5\columnwidth]{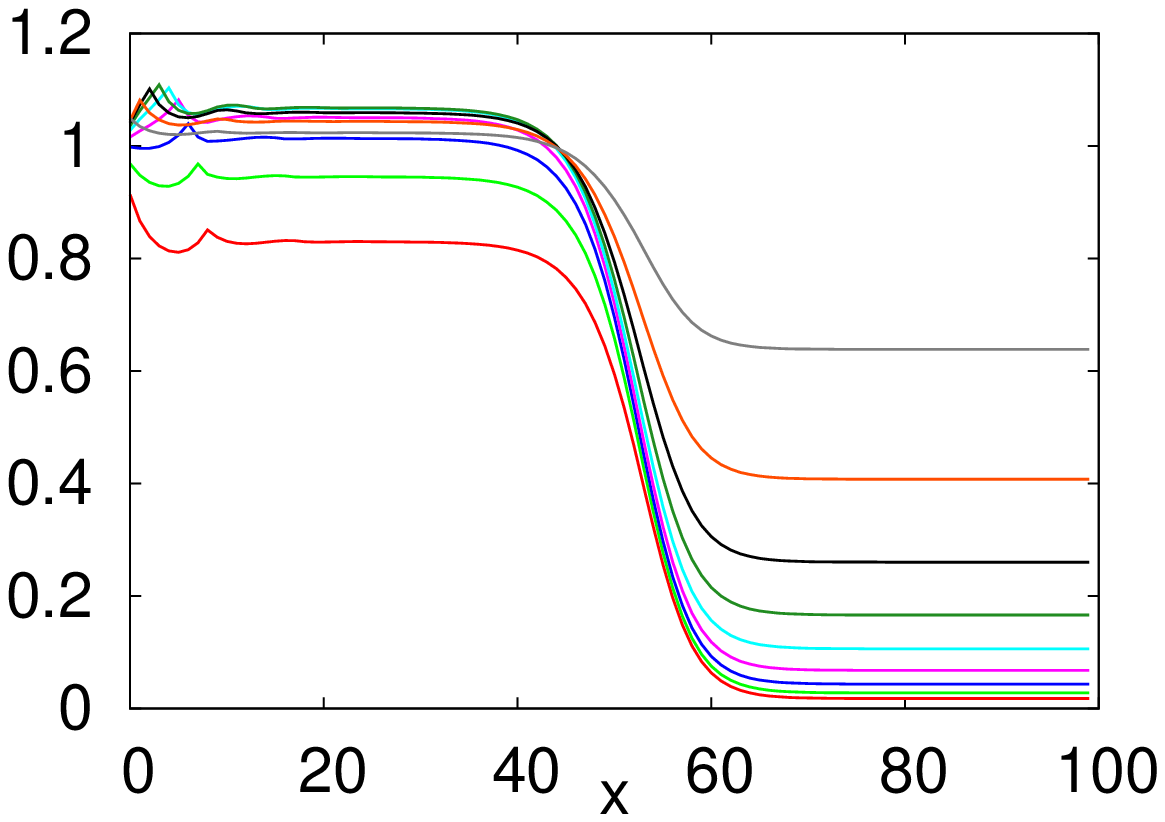}}
\subfloat{\includegraphics[width=0.5\columnwidth]{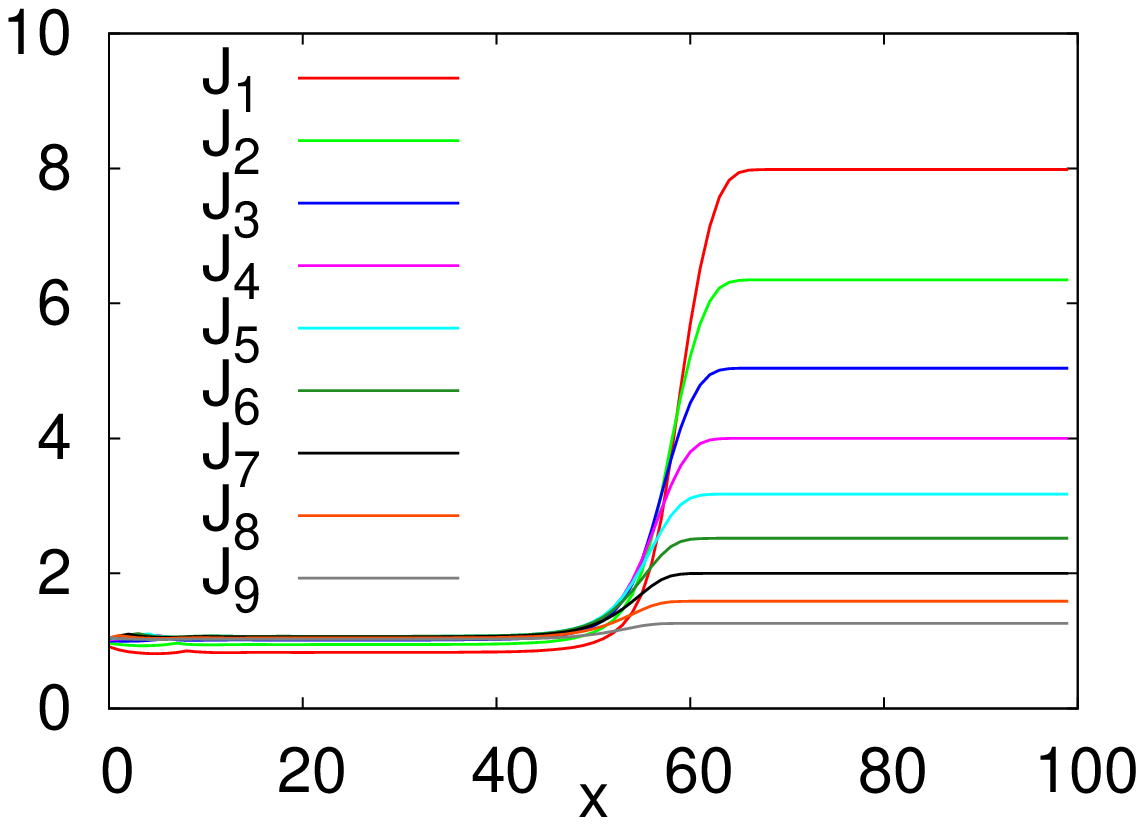}}
\end{center}
\caption{(color online) Renormalized couplings $J'$ in a FM system
  with $n=10$ levels versus the number of RG steps, for temperatures
  slightly bigger (left) and smaller (right) than $T_c$.}
\label{coupling}
\end{figure}

Applying the previously described procedure, the flux of couplings and
correlations can be followed, see Fig.~\ref{coupling}.  If $T\simeq
T_c$, renormalized couplings stay for a while close to the critical
fixed point (FP), and then go towards the high temperature (HT) FP if
$T>T_c$ or the low temperature (LT) FP if $T<T_c$.  We estimate the
critical temperature as the temperature dividing the flows towards the
two different FP.  Please note that the HT and LT fixed points are not
characterized by the usual $J=0$ and $J=\infty$ coupling values: the
reason is that in step 3 of our procedure we put new coupling of
original intensity. Nonetheless couplings flows clearly differentiate
HT and LT behaviors.

\begin{figure}[t]
\begin{center}	
\includegraphics[width=0.8\columnwidth]{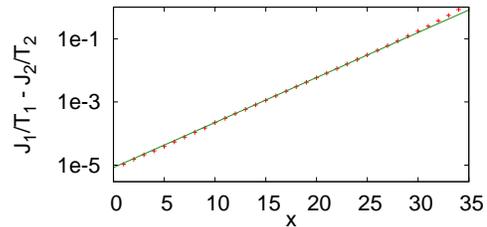}
\end{center}
\caption{\label{nuD3} Difference between the renormalized couplings at
  $T_1=0.84571$ and $T_2=0.845716$ in the FM HM with $n=13$ levels and
  $D\simeq3$.  The fit estimates the $\nu$ exponent.}
\end{figure}

In order to extract critical exponents from the RG equations, we focus
on the early regime, when the coupling flows leave the critical
FP. From the Wilson relation an equation relating renormalized
coupling after $x$ RG steps can be obtained:
\begin{equation*}
\frac{J_1(x)}{T_1} - \frac{J_2(x)}{T_2} = \((\frac{1}{T_1} - \frac{1}{T_2}\)) b^{\frac{x}{\nu}}\;,
\end{equation*}
where $b=2$ is the scaling factor in our case. Thus, the $\nu$
exponent can be estimated from a fit like the one in Fig.~\ref{nuD3}.
The values obtained for the critical temperature and the critical
exponents in $D\simeq3$ are well comparable with the known ones: $T_c$
extrapolates to $0.8478(1)$ in the large $n$ limit and the measured
$\nu$ exponent is $2.076(6)$ for $n=13$ (although the extrapolation of
$\nu$ to the $n\to\infty$ limit is difficult due to strong finite size
effects, e.g. $\nu=2.79(12)$ for $n=4$).  Moreover we have checked
that our numerical RG recovers the right bounds on $c$, namely $T_c
\to 0$ for $c_L=1/4$ and $T_c \to \infty$ for $c_{\infty}=1/2$.

We consider now the DFM. The Hamiltonian of the model is always the
one in Eq.~(\ref{eq:Hindip}), but the couplings at level $k$ are
independent random variables extracted from the distribution
\begin{equation*}
P_k(J) = p_k\,\delta(J-J_k) + (1-p_k)\,\delta(J)\;.
\end{equation*}
At the beginning $p_k=p$ and $J_k=1$ for any $k$, while under the RG
they will differentiate.  The number of parameters to be determined in
the ERG is $2(n-1)$, and we use the following observables, with
$k\in\{1,\ldots,n-1\}$, to fix them:
\begin{equation*}
\overline{\((\frac{\langle m_{L_k} m_{R_k}\rangle}{\langle m_{L_k} m_{L_k}\rangle}\))}\quad,
\quad\overline{\((\frac{\langle m_{L_k} m_{R_k}\rangle}{\langle m_{L_k} m_{L_k}\rangle}\))^2}\;. 
\end{equation*}
\begin{figure}[t]
\begin{center}	
\includegraphics[width=0.8\columnwidth]{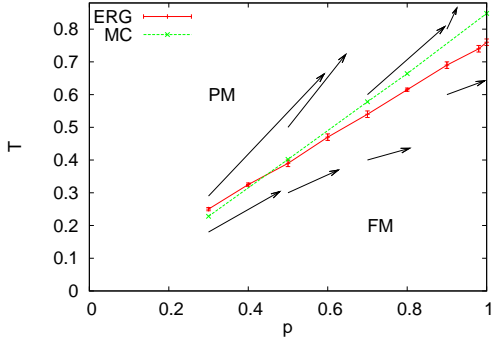}
\end{center}
\caption{\label{dil} Phase diagram in the $p-T$ plane for the DFM at
  $D\simeq3$ as obtained by the ERG and by MC simulations.  Arrows
  represent the first iteration of the ERG.}
\end{figure}
Applying the same procedure as for the pure model, we are able to draw
a flow diagram in the $p$-$T$ plane for $D\simeq3$ and determine the
critical line (see Fig.~\ref{dil}). The validity of the phase diagram
found with the ERG is confirmed by a set of MC simulations (explained
later in the text) whose $T_c$ estimates are also shown in
Fig.~\ref{dil}.  The only disappointment about this phase diagram is
that we do not find an unstable FP along the critical line as expected
for a $D=3$ SR model \cite{diluted3D}.  However this can be explained
by noticing that the $\alpha$ exponent of this model is very small,
$\alpha=0.051288905$, and so the crossover from the pure behavior can
be extremely long.

\begin{figure}[t]
\begin{center}
\includegraphics[width=\columnwidth]{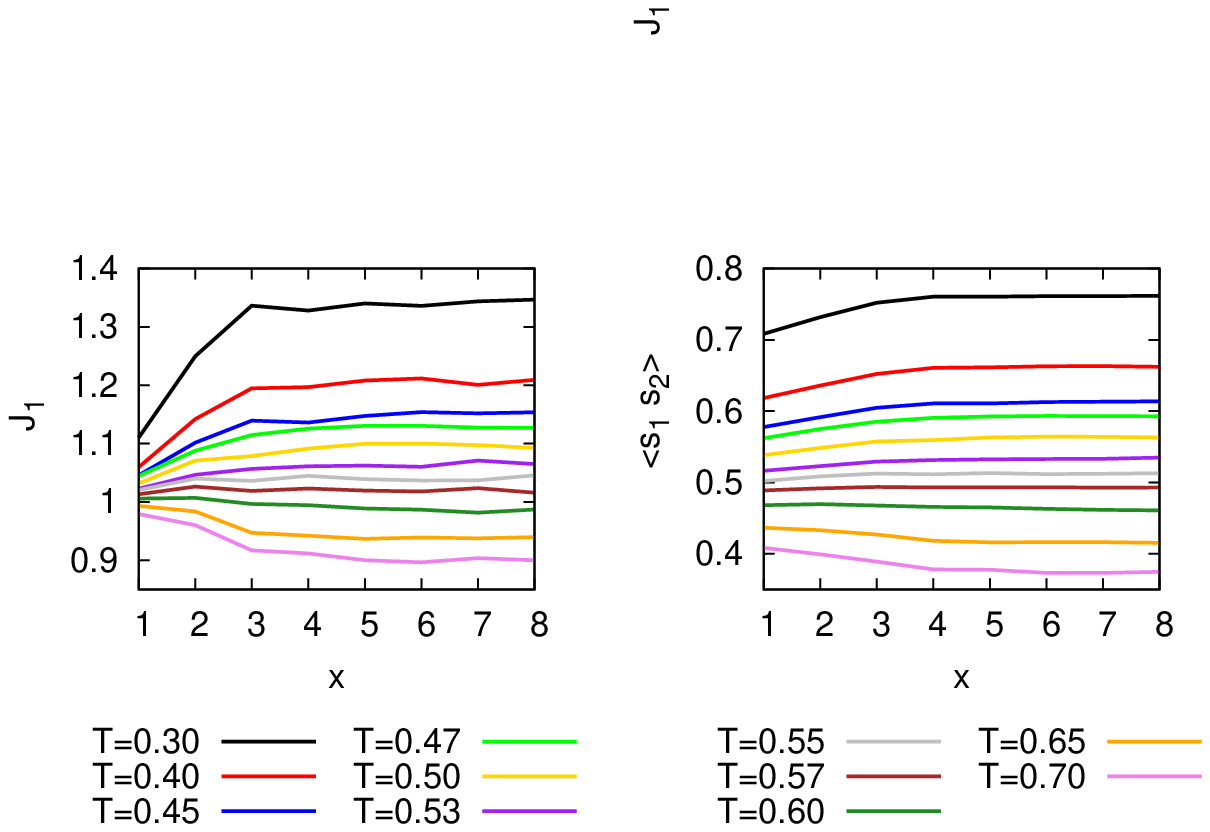}
\end{center}
\caption{(color online) Variances of renormalized couplings (left) and
  of SG correlations (right) at the lower level ($k=1$) in a SG system
  with $n=4$ levels and $D\simeq3$ versus the number of
  renormalization steps $x$, for many temperatures. We locate the
  critical temperature at $T_{SG}=0.58(1)$.}
\label{d3gauss}
\end{figure}

Finally we study the SG version. The Hamiltonian is always the one in
Eq.~(\ref{eq:Hindip}), and the couplings at level $k$ are distributed
with a Gaussian law of zero mean and variance $\sigma_k^2$ (at the
beginning $\sigma_k^2=1$ for any $k$).  The assumptions that the
renormalized couplings are independent and normally distributed could
be released by adding extra terms in the coupling distributions
\cite{ParisiPetronzio}, but we leave these generalizations for future
works.  In the SG case the observables used to fix the $n-1$ variances
are normalized SG correlations at different levels:
\begin{equation*}
\overline{\< O_k \>} = \overline{\frac{\sum_{i\in L_k,\;j \in R_k} \<s_i s_j\>^2 }{\sqrt{ \sum_{i,j \in L_k}\<s_i s_j\>^2   \sum_{i,j \in R_k}\<s_i s_j\>^2 }}}\;.
\end{equation*}
Because of the computational costs we use $n \le 4$, so the early
regime leaving the critical FP is rather short, and the stationary
regime is soon reached (with respect to the FM case).  This effect is
also enhanced by the disorder: indeed, even exactly at criticality,
the SG ensemble contains many samples which are not critical, and the
couplings of these samples flow away from critical values very fast.
So, it seems unavoidable that disorder increases the instability of
critical FP and consequently the uncertainty on the estimates of
critical exponents.  Nonetheless we can distinguish two temperature
regions separated by a critical temperature $T_{SG}$ (see
Fig.~\ref{d3gauss} for an effective dimension $D\simeq3$ and $T_{SG} =
0.58(1)$), such that above $T_{SG}$ correlations and couplings decay
towards zero, while below $T_{SG}$ correlations and couplings
variances grow, suggesting that the system is in a SG phase.  In
Fig.~\ref{d3gauss} we have plotted only couplings and correlations
variances measured at the lowest level $k=1$, but (as in the FM, see
Fig.~\ref{coupling}) the renormalized variances at the other levels
are related to those at $k=1$: for example, $\sigma_2 > \sigma_1$ if
$T > T_{SG}$ and $\sigma_2 < \sigma_1$ if $T < T_{SG}$ (remember that
parameters at the lowest level are those which are less influenced by
the coupling of the original intensity at level $n$, because they have
been renormalized more times).

\begin{figure}[t]
\begin{center}	
\subfloat{\includegraphics[width=0.5\columnwidth]{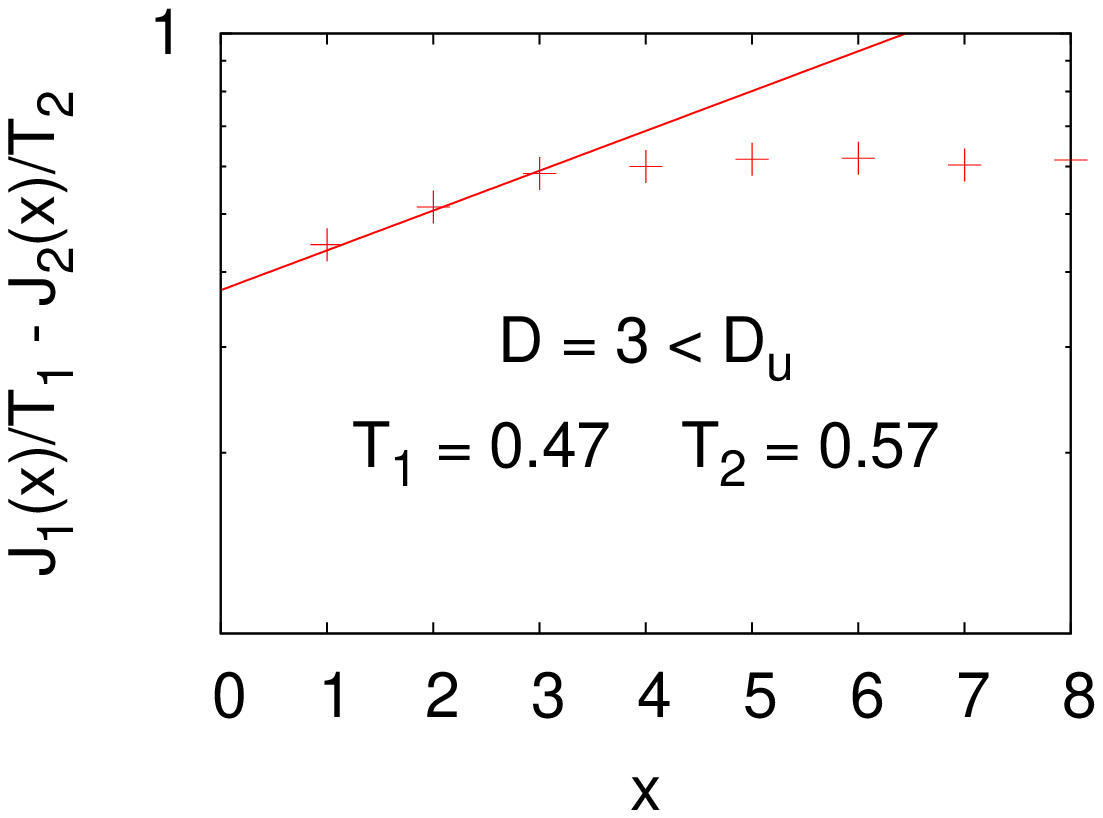}}
\subfloat{\includegraphics[width=0.5\columnwidth]{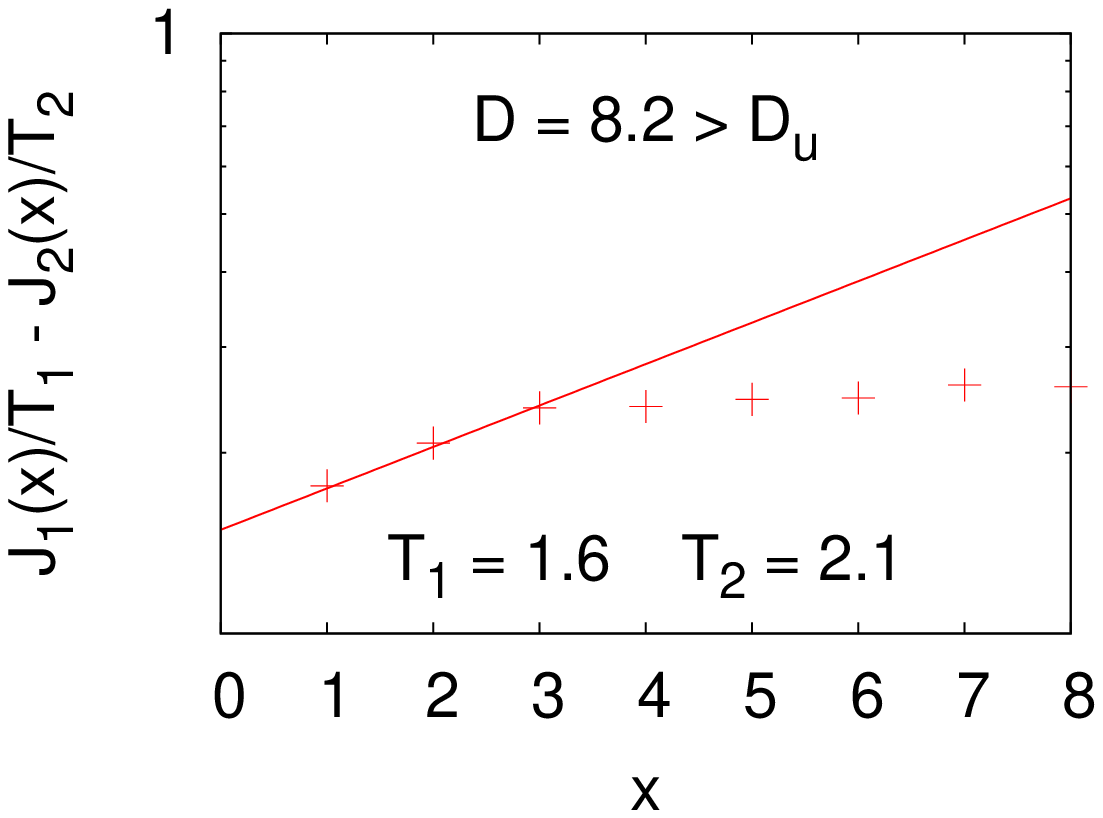}}
\end{center}
\caption{\label{nu_gauss} Difference between the renormalization flux
  of the couplings at two different temperatures in a semi-log scale
  for $D\simeq3<D_u$ (left) and $D\simeq8.2>D_u$ (right). The first
  part, not affected by finite size effects, has been used to extract
  the $\nu$ exponent, through a power law fit.}
\end{figure}

\begin{figure}[t]
\begin{center}	
\includegraphics[width=\columnwidth]{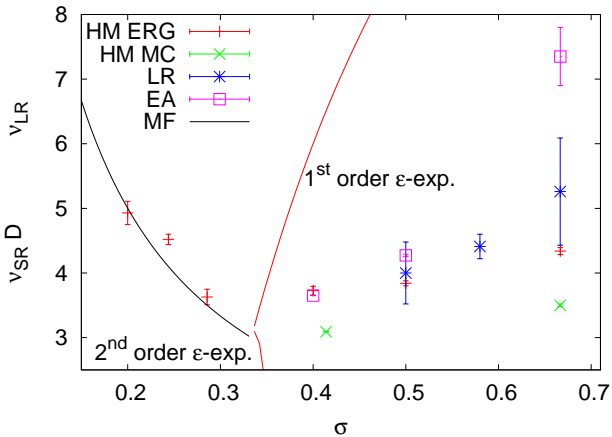}
\end{center}
\caption{(color online) The estimation of the $\nu$ exponent from the
  ERG for different values of $c=2^{-(1+\sigma)/2}$. The curve for
  $\sigma \le 1/3$ is the MF prediction. We also show results from the
  $\varepsilon$-expansion and from MC simulations of the HM. For
  comparison, $\nu$ values for the power-law LR model and $D \nu$
  values for the EA model in $D=3,4,5$ are added.}
\label{nu_SG}
\end{figure}

Also in the SG case we are able to estimate the $\nu$ exponent from
the flux of the couplings at early times.  The procedure used is the
same of that in the FM case and typical fits are shown in
Fig.~\ref{nu_gauss} for $D\simeq8.2$ in the mean-field region and for
$D\simeq3$ below the upper critical dimension.  We obtain $\nu =
4.15(10)$ in $D\simeq8.2$ and $\nu = 4.34(6)$ in $D\simeq3$.  In
Fig.~\ref{nu_SG} we report (with label HM ERG) the estimates of $\nu$
as a function of the exponent $\sigma$ defined as
$c=2^{-(1+\sigma)/2}$: we see that in the mean-field region
($\sigma<\frac{1}{3}$) results are compatible with the expected
mean-field behavior ($\nu=D/2$ that corresponds to
$\nu=\frac{1}{\sigma}$) and, more interestingly, the critical exponent
$\nu$ has a minimum around the upper critical dimension $D_U=6$,
$\sigma_U=1/3$, as confirmed by our MC simulations (label HM MC in
Fig.\ref{nu_SG}) and those in Ref.~\cite{HEAM}.  This minimum was not
observed in previous RG studies \cite{RSRGCastellana}, while it is
present in SR models: in Fig.~\ref{nu_SG} we report the $D \nu_{SR}$
estimates for the EA model in $D=3,4,5$
\cite{D3Pellissetto,sigmad,D5Aharony} following the relation
introduced previously: $D=\frac{2}{\sigma}$.  The same non-monotonic
behavior for the $\nu$ exponent has been also seen in a $1d$ SG model
with long range (LR) power law decaying interactions ($J_{ij}\propto
|r_{ij}|^{-(1+\sigma)/2}$) \cite{Levy,sigmad}, also shown in
Fig. \ref{nu_SG}.

\section{Monte Carlo simulations}

In order to check critical temperatures and the critical exponent
$\nu$ for the SG version of HM in the non-mean-field region, we have
run MC simulations at several values of $c$. Couplings are extracted
from a Gaussian and from a binary $\pm J$ distributions.  We have used
the parallel tempering algorithm \cite{paralleltemp}, running
simultaneously at $20$ different temperatures.  Two replicas have been
simulated in parallel, to measure the overlap between them.
Equilibration has been checked by the standard method of observing the
convergence of the measured observables (e.g.\ energy and overlap
moments) to their asymptotic values.  We have found that the
equilibration time is $\tau\simeq10^5-10^6$ MC steps for the largest
sizes at smaller temperatures and we have acquired data for
$5\cdot\tau$ MC steps.  For the model with Gaussian couplings we have
used $n=6,7,8,9,10$, while for the one with $\pm J$ couplings, that is
easier to simulate, $n=6,7,8,9,10,11$.  Averages were performed over
400 samples for the larger systems up to 2000 samples for the smaller
ones.

We have run also standard MC simulations (with non parallel tempering)
for the DFM at several bond concentrations ($p=0.3,0.5,0.7,0.8$).

Critical temperatures have been estimated from the crossing points of
the scale-invariant observables: these crossing points should approach
the critical temperature for large sizes.  In particular we study the
dimensionless Binder parameter \cite{Binder}, defined as
$B=\frac{1}{2}\[[3-\frac{\langle m^4 \rangle}{\langle m^2
  \rangle^2}\]]$ for the DFM and
$B=\frac{1}{2}\[[3-\frac{\overline{\langle q^4
    \rangle}}{\overline{\langle q^2 \rangle}^2}\]]$ for the SG.  We
can construct also another scale-invariant observable from the
susceptibilities, $\chi_{FM}=m_2L$ and $\chi_{SG}=q_2L$, knowing
analytically its dimension $2-\eta$.  In fact the $\eta$ exponent is
not renormalized in LR systems.  Defining $c=2^{-(1+\sigma)}$ for the
FM and $c=2^{-(1+\sigma)/2}$ for the SG, $\eta=2-\sigma$ in the whole
region $c_L<c<c_{\infty}$.  Thus if we divide $\chi$ by its dimension,
the resulting $m_2L^{-1+\eta}$ or $q_2L^{-1+\eta}$ should remain
finite at the critical temperature. We can apply the same argument to
the fourth moment susceptibility, obtaining another scale-invariant
observable $m_4L^{-2+2\eta}$ for DFM and $q_4L^{-2+2\eta}$ for SG.

\begin{figure}[t]
\begin{center}	
\includegraphics[width=\columnwidth]{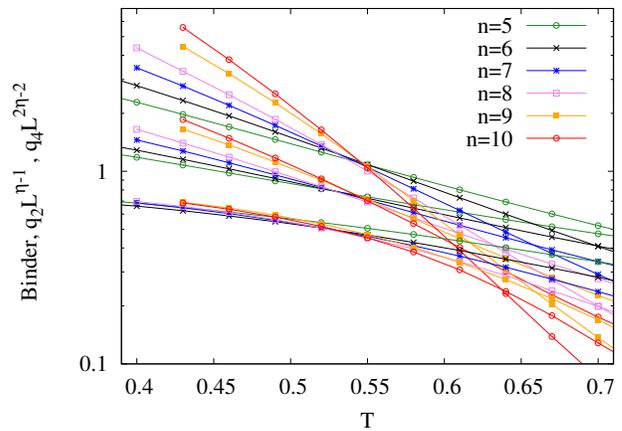}
\end{center}
\caption{(color online) Scale-invariant observables, $B$,
  $q_2L^{-1+\eta}$ and $q_4L^{-2+2\eta}$ for different sizes $L=2^n$
  as a function of the temperature for a SG HM with
  $c=2^{-\frac{5}{6}}$ and a Gaussian distribution of the couplings.
  The crossing points locate the critical temperature $T_c=0.55(1)$}
\label{Tc}
\end{figure}

In Fig. \ref{Tc}, we plot the three scale-invariant observables, $B$,
$q_2L^{-1+\eta}$ and $q_4L^{-2+2\eta}$ for different sizes $L=2^n$ as
a function of the temperature for a SG HM with $c=2^{-\frac{5}{6}}$
and a Gaussian distribution of the couplings.  The crosses of the
curves should approach the critical temperature, that we estimate to
be $T_c=0.55(1)$.  Unfortunately the data do not allow us to estimate
the correction to scaling exponent $\omega$.  Nonetheless we can say
that the Binder parameter is the observable that has the largest
finite size effects.  The critical temperature is compatible with the
one obtained from the ERG $T_c=0.58(1)$.  Please consider that the
estimate from ERG is obtained with a small number of levels $n=4$ (and
in a much faster time).  In Fig.~\ref{dil} we report the critical
temperatures of the DFM obtained by an analysis very similar to the
one just described.

To measure the critical exponent $\nu$ we have used the more accurate
data for the model with $\pm J$ interactions, that should be in the
same universality class.  We use the same observables previously
described to determine the critical temperature that we estimate as
$T_c=0.545(10)$ for $c=2^{-\frac{5}{6}}$.  Then we look at the values
of $d(q_2L^{-1+\eta})/dT$ at $T_c$. The dimension of this observable
is $1/\nu$ and its finite size scaling form at leading order around
$T_c$ is described by
$$
d(q_2L^{-1+\eta})/dT=L^{1/\nu}F(L^{1/\nu}(T-T_c))\;.
$$
\begin{figure}[t]
\begin{center}	
\includegraphics[width=0.9\columnwidth]{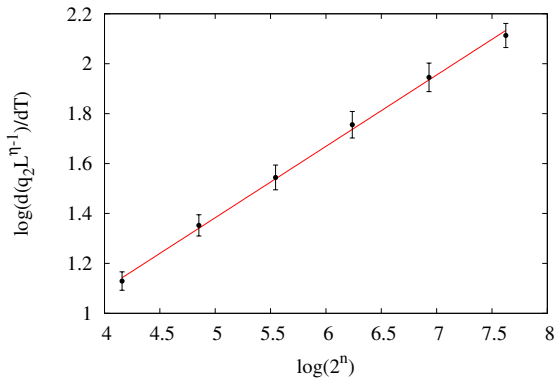}
\end{center}
\caption{$d(q_2L^{-1+\eta})/dT$ at $T_c$ for different sizes as a
  function of the size of the system in a log-log scale. The straight
  line is the best fit to extract the exponent $\nu$. The system has
  $\pm J$ interactions and $c=2^{-\frac{5}{6}}$, $D\simeq3$.}
\label{nuMCbinary}
\end{figure}
In Fig. \ref{nuMCbinary} the values for $d(q_2L^{-1+\eta})/dT$ at
$T_c=0.545$ for different sizes are plotted as a function of the size
of the system in a log-log scale. We have chosen this particular
observable because we have seen that it has the smallest finite size
effects.  In this scale $1/\nu$ can be extracted via a linear fit as
the angular coefficient. The straight line is the best fit that leads
to $\nu=3.50\pm0.02$.  This result should be compared with the ERG
estimate $\nu=4.34(6)$. They are rather different. However we know
that for the estimate of $\nu$ exponent with the ERG method we have
very large finite size effects, especially in the non mean field
region (indeed we have an error of $30\%$ for $n=4$ already for the
ferromagnetic case).  Nevertheless the important result is that the
non monotonic behaviour of $\nu$ with $c$ is confirmed by the MC
simulations.

\begin{figure}[t]
\begin{center}	
\includegraphics[width=\columnwidth]{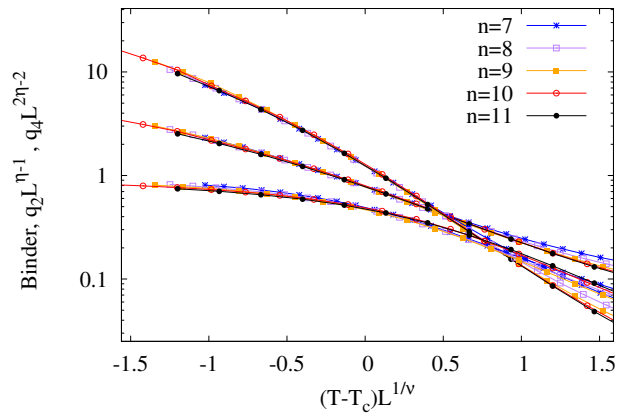}
\end{center}
\caption{(color online) $B$, $q_2L^{-1+\eta}$ and $q_4L^{-2+2\eta}$
  for different sizes as a function of $L^{1/\nu}(T-T_c)$. Curves at
  different sizes collapse quite well. The system has $\pm J$
  interactions and $c=2^{-\frac{5}{6}}$, $D\simeq 3$.}
\label{MC_scaling}
\end{figure}

In Fig. \ref{MC_scaling} we have plotted the three scale-invariant
observables $B$, $q_2L^{-1+\eta}$ and $q_4L^{-2+2\eta}$ for different
sizes as a function of $L^{1/\nu}(T-T_c)$, with $\nu=3.5$ and
$T_c=0.545$. We can see that curves at different sizes collapse quite
well.

\section{Conclusions}

Concluding, we have developed a semi-analytical real space RG method
that can be used for disordered systems.  The method has been applied
to the hierarchical model and is able to find a SG transition also for
effective dimension $D\simeq3$ in the non-mean field region.  The
reliability of the method has been tested comparing the values of
critical temperatures and critical exponents with those obtained in MC
simulations.  The agreement for the critical temperatures is optimal
for all the version we have studied and the ERG method allows to
compute them in a time much smaller than MC simulations.  The critical
exponents obtained with the ERG method are more affected by finite
size effects than critical temperatures.  In fact the error in the ERG
estimate for the SG exponents is around 20\%. However such a large
error is not unexpected, since we find it also in the ferromagnetic
case, where the correct exponent is recovered only in $n\to\infty$
limit.

The ERG method is able to reproduce the correct behavior of the $\nu$
exponent, that shows a minimum at the upper critical value of $c=c_U$
(or $\sigma=\sigma_U$): such a behavior is predicted by the
$\epsilon$-expansion of the SG version \cite{eps_exp} and found in our
MC simulations.  This accurate prediction by the ERG make us confident
that the method is reliable.  The present results by the ERG solve an
apparent inconsistency problem between the $\epsilon$-expansion and
another real space RG approach recently proposed in
Ref.~\cite{RSRGCastellana}, where the $\nu D$ exponent was found to
decrease linearly with $D$, with no minimum at all around the upper
critical dimension.  In this work we have shown that such a minimum in
$\nu D$ exists if a better RG transformation is used.  In particular
we have shown that the right way to do renormalization when disorder
is present is to consider RG on ensemble of systems and not on single
samples.

This statement can be confirmed also from a comparison with the method
of analysis used in MC simulations.  When exponents are determined
using finite size scaling (FSS), we look at the crossing point of some
dimensionless quantities (like the Binder parameter) averaged over
disorder for different sizes.  Indeed FSS is an approximate RG where
only a parameter, the temperature, is varied. In this perspective, our
RG is a generalization of FSS, where we vary more parameters, and for
this reason we can look at more observables. However, like in FSS, the
useful observables are those averaged over the disorder.


\begin{thebibliography}{99}
\bibitem{cDrelations} We are aware that more accurate $c-D$ relations
  can be defined, as discussed exhaustively in Ref.~\cite{sigmad}.
  However the scope of our comparison with short range models in the
  present work is only qualitative and does not pretend to be
  quantitative.  For this reason the simplest relation is sufficient
  for our purpose.
\end{thebibliography}
\end{document}